# Artificial versus Biological Intelligence in the Cosmos:

# Clues from a Stochastic Analysis of the Drake Equation

Alex De Visscher

Department of Chemical and Materials Engineering, Gina Cody School of Engineering and Computer Science, Concordia University, Montreal, Quebec, Canada. Tel.: +1-514-848-2424 ext. 3488. E-mail: alex.devisscher@concordia.ca

**Abstract**

The Drake equation has been used many times to estimate the number of observable civilizations in the Galaxy. However, the uncertainty of the outcome is so great that any individual result is of limited use, as predictions can range from a handful of observable civilizations in the observable universe to tens of millions per Milky Way-sized galaxy. A statistical investigation shows that the Drake equation, despite its uncertainties, delivers robust predictions of the likelihood that the prevalent form of intelligence in the universe is artificial rather than biological. The likelihood of artificial intelligence far exceeds the likelihood of biological intelligence in all cases investigated. This conclusion is contingent upon a limited number of plausible assumptions. The significance of this outcome in explaining the Fermi paradox is discussed.

**Introduction**

Thousands of exoplanets have been discovered in the last two decades. This has spurred an increased interest in exobiology (e.g., Schneider, 2016). It has been argued that our current concepts on extraterrestrial intelligence should be reconsidered in the light of recent advances in the field of artificial intelligence. It is now conceivable that the dominant form of intelligence in the universe may be artificial rather than biological (Shostak, 2018; Gale et al., 2020). While the emergence of artificial intelligence represents an additional filter, such an intelligence may be more long-lived than its creators, offsetting the effect of the additional filter.

The Drake equation (Drake, 1965) can be used to evaluate the abundance of detectable extraterrestrial intelligence in the Galaxy. While variants exist, the most commonly used form of the equation is as follows:

$$N = R^* f_p n_e f_l f_i f_c L \qquad (1)$$

where $R^*$ is the rate of star formation in the Galaxy (yr$^{-1}$), $f_p$ is the fraction of stars with planets, $n_e$ is the average number of earth-like planets that are potentially habitable, per star, $f_l$ is the fraction of habitable planets where complex life develops, $f_i$ is the fraction of life-bearing planets that develop intelligence, $f_c$ is the faction of intelligent life-bearing planets where observable technology develops, and $L$ is the mean duration of these technological civilizations.

Seager (2018) developed a modified Drake equation to guide searches for biosignatures from observable planets, and concluded that the number of observable biosignatures with current technology is as low as 1-4, depending on the technology, even assuming very optimistic probabilities of life occurring on habitable planets.

The result of the Drake equation is generally understood to refer to biological intelligence, although this is not usually stated explicitly. However, there is no reason why it could not be used to evaluate the prevalence of artificial intelligences. However, predictions made with the Drake equation differ by as many as eight orders of magnitude (Sandberg et al., 2018), which means that a simple comparison of two predictions (biological vs. artificial intelligence) with the equation is meaningless without context.

A first attempt to get a better grip on the variables in the Drake equation based on Monte Carlo simulations was made by Forgan (2009). The number of advanced civilizations in the Milky Way estimated in this study ranged from 360 to 38,000, depending on the assumptions made.

An attempt to evaluate the uncertainty of the Drake equation with a statistical argument is by Maccone (2010), who found that the number of observable extraterrestrial intelligences in the Galaxy runs in the thousands, but with a standard deviation exceeding the mean. Glade et al. (2012) developed a stochastic model based on the Drake equation with the purpose of making the estimation time-dependent. More recent statistical approaches are by Engler and von Wehrden (2019), and by Bloetscher (2019). Estimates of the latter two are widely divergent, albeit by very different methods: between 7 and 300 technological species over the entire life span of the Milky Way to date (Engler and von Wehrden, 2019), and between 2 and 250 intelligent civilizations in the Milky Way at any given time (Bloetscher, 2019). Sandberg et al. (2018) conducted a Monte Carlo simulation based on a set of variables used in the literature for the Drake equation. They concluded that the proportion of possible model variants leading to the conclusion that we are alone in the Milky Way is about 30 %. A second simulation that was not constrained to parameter values found in the literature led these authors to conclude that the likelihood of an empty galaxy exceeds 30 %. In the studies of Engler and von Wehrden (2019), Bloetscher (2019), and Sandberg et al. (2018), the absence of observed technological signals is attributed to the sparseness of technological life in the Milky Way.

The purpose of this study is to evaluate the likelihood that the universe is dominated by artificial rather than biological intelligence. To that effect, Monte Carlo simulations are conducted with two linked versions of the Drake equation: one to estimate the number of observable extraterrestrial biological intelligences, and one to estimate the number of observable extraterrestrial artificial intelligences.

**Methodology**

Equation (1) is used as the basis of a Monte Carlo calculation. For each of the parameters in Eq. (1), a probability density function is assumed. A large number of samples are taken from each distribution, and used in Eq. (1) to obtain a sample of the probability distribution of the number of observable intelligences in the Milky Way. Following Sandberg et al. (2018)'s second simulation, we use a log-uniform distribution for the variables $R^*$, $f_p$, $n_e$, $f_i$, and $f_c$. The assumed ranges are those of Sandberg et al. (2018): 1-100 for $R^*$, 0.1-1 for $f_p$, 0.1-1 for $n_e$, and 0.001-1 for $f_i$.

For $f_c$, we distinguish between $f_{c,b}$, the probability that an intelligent biological species develops the ability to communicate over interstellar space, and $f_{c,AI}$, the probability that an intelligent biological species develops an artificial intelligence capable of communicating over interstellar space. For the former, we adopt Sandberg et al. (2018)'s range of 0.01-1, whereas for the latter, we assume a range of 0.0001-1. As a justification of this range, it is assumed that the development of an artificial intelligence represents an additional filter with the same selectivity as the filter of an intelligent species developing the ability to communicate across interstellar space. If $f_{c,AI}$ results from two filters with selectivity $f_{c,b}$, in series, the following relationship applies:

$$f_{c,AI} = f_{c,b}^2 \qquad (2)$$

The variables $f_l$ and $L$ represent the greatest uncertainty, and require further reasoning. For $f_l$, Sandberg et al. (2018) recommend an equation of the following form:

$$f_l = 1 - \exp(-k) \qquad (3)$$

where $k$ is a log-normally distributed variable. Sandberg et al. (2018) recommend an average value of $k$ of 1 and a standard deviation of 50 orders of magnitude for their second simulation. The latter was motivated by the observation that estimations of the probability of emerging life spans 200 orders of magnitude. This extreme range is informed in part by estimates of the probability of randomly synthesizing RNA polymers of the correct structure and of sufficient length to self-replicate. Studies of this nature argue that an inflationary universe is needed to explain the emergence of life, and suggest that we are not only alone in the universe, but alone in a multiverse many orders of magnitude larger than the observable part of the universe (e.g., Totani, 2019). However, Spiegel and Turner (2012) pointed out that life emerged on earth within a few hundred million years after the planet cooled down to a temperature that can support life. This could indicate that the emergence of life is easier than the emergence of intelligence, which took billions of years. Spiegel and Turner (2012) point out that this argument is inconclusive, though.

Models that require a multiverse vastly larger than the observable universe are problematic because they are untestable outside the parameter space corresponding with the size of the visible universe. For that reason, a much more narrow range was considered here. This is an *a priori* assumption (Mix, 2018) made for pragmatic reasons. I maintained eq. (3) with a lognormal distribution for $k$. The mean and standard deviation were adjusted so that the distribution of $N$ closely matches the distribution of $N$ resulting from Sandberg et al. (1918)'s *first* calculation (i.e., the calculation based on a sampling of parameters proposed in the literature rather than parameters drawn from distributions), with the exception of the low-probability tail. A good agreement with my "optimistic scenario" (see below) was obtained when the log-normal distribution of $k$ has variables $\mu = -2$ and $\sigma = 7.5$. This leads to a median value of $f_l$ of 0.126 (theoretical value: 0.127) and an average value of 0.425. On the other hand, the 10$^{th}$ percentile of $f_l$ is $9.1 \times 10^{-6}$.

For the remaining parameter, $L$, a couple of variants were explored. Sandberg et al. (1918)'s calculation involved a log-uniform distribution from 100 to $10^{10}$ years. I adopted this distribution for the artificial intelligence in the base calculation. This leads to a median duration of an intelligent civilization of a million years. This estimate may be optimistic in the case of a biological intelligence, given the ways an intelligent civilization can destroy itself (e.g., biological, Sotos (2019)). For that reason, a power law in log scale is chosen for the duration of biological intelligences in the base calculation, so that the median value of $L$ is 1000 years, while maintaining the 100-$10^{10}$ years range. This corresponds with a power-law index of -2/3.

A second, more optimistic scenario is simulated, where $L$ for biological intelligences has the same log-uniform distribution in the 100-$10^{10}$ years range as for the artificial intelligence. This scenario probes the probability of an artificial intelligence prevailing in spite of the additional filter, without the benefit of a longer life span probability distribution.

A third scenario is the base scenario for biological intelligence, but with a long-life artificial intelligence lifetime probability distribution. To this effect, the following equation is used:

$$L = L_{max}\left(1 - \exp(-k_L)\right) \qquad (4)$$

where $L_{max} = 10^{10}$ years, and $k_L$ is a lognormally distributed variable with distribution parameters $\mu$ and $\sigma$ chosen so as to obtain a median value of $L$ of $10^8$ years, and a mean value of $10^9$ years. This is realized when $\mu = -4.6052$ and $\sigma = 2.865$.

For each calculation, the simulation is run for 100,000 iterations multiple times, and the key statistical properties are compared for robustness of the obtained results. The results were identical to within a few tenths of a percent in all cases, except for percentile values, including medians, where the variation was up to a few percent. Whenever statistical properties are reported, they are the mean of at least five simulations with 100,000 iterations, or at least one simulation with one million iterations.

**Results and Discussion**

*Distributions of N*

Figure 1 shows the cumulative probability density distributions of the number of biological intelligences in the Milky Way, and the number of artificial intelligences in the Milky Way, in the base case.

The distributions span nearly 20 orders of magnitude. The number of detectable biological intelligences ranges from $3.15 \times 10^{-9}$ at the 1$^{st}$ percentile, to $3.74 \times 10^7$ at the 99$^{th}$ percentile, with a median value of 0.460 detectable biological intelligences in the Milky Way. To put this into perspective, the 1$^{st}$ percentile corresponds with on the order of 30 detectable biological intelligences in the universe, whereas the 99$^{th}$ percentile corresponds with about one detectable biological intelligence every 10,000 stars.

The number of detectable artificial intelligences is roughly equal to the number of detectable biological intelligences. The longer expected time of existence of a detectable artificial intelligence roughly compensates for the additional filter needed to create the artificial intelligence. The first percentile of $N$ is $1.04 \times 10^{-9}$, the median is 0.679, and the 99th percentile is $2.52 \times 10^6$.

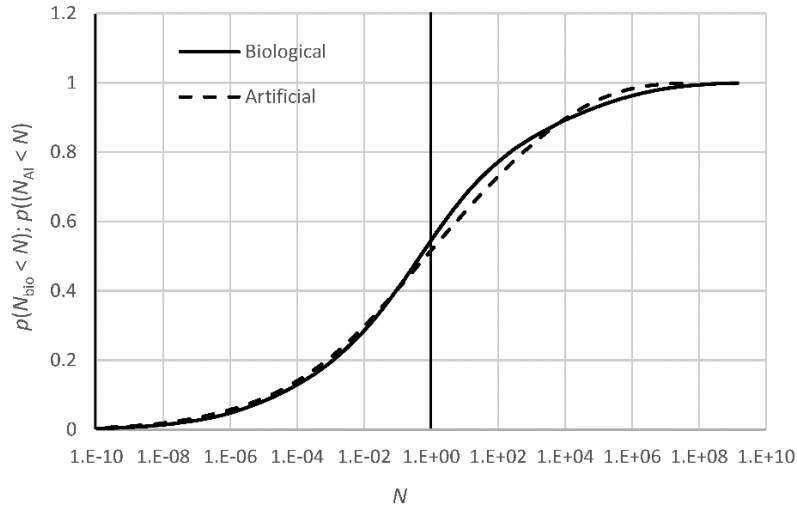

Figure 1. Cumulative distribution of probabilities of the number of detectable intelligences in the Milky way, biological (solid line) and artificial (dashed line), base case. Vertical line in the middle represents threshold for no (other) intelligence in the Milky Way; vertical line on the left represents threshold for no (other) intelligence in the universe.

With the median life span of biological life set to 1000 years, it was impossible to reproduce the cumulative distribution of $N$ of Sandberg et al (2018)'s first simulation for probabilities above 30 %, even when it was assumed that the probably of life emerging is unity. This is why the choice of parameters of the distribution of $f_l$ was set based on a simulation with the same distribution of $L$ as Sandberg et al. (2018). This represents a more optimistic scenario for the survival of intelligent life, with a median value of $L$ of one million years. The cumulative distribution for this scenario is shown in Figure 2.

As could be expected, the distribution of the number of detectable biological intelligences has moved to higher values, due to the longer survival times. The spread of the distribution is roughly the same, but the distribution has moved by approximately two orders of magnitude. The 1st percentile of $N$ is now $1.10 \times 10^{-7}$, the median is 68.3, and the 99th percentile is $2.51 \times 10^8$. The 99th percentile corresponds with nearly one detectable biological intelligence every thousand stars.

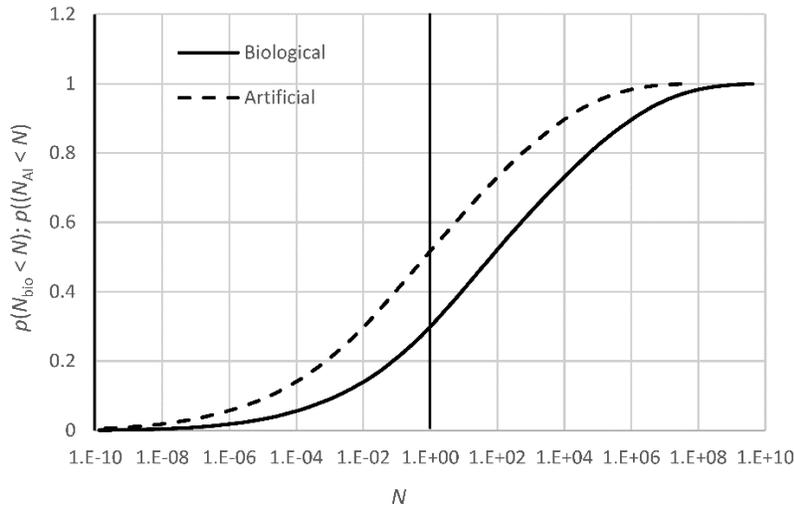

Figure 2. Cumulative distribution of probabilities of the number of detectable intelligences in the Milky Way, biological (solid line) and artificial (dashed line), "optimistic", long biological survival time scenario. Vertical line in the middle represents threshold for no (other) intelligence in the Milky Way; vertical line on the left represents threshold for no (other) intelligence in the universe.

In the third scenario, we assume a mean survival time of detectable artificial intelligence of a billion years, and a median survival time of 100 million years. For comparison, for detectable biological intelligences, the base distribution with a median survival time of 1000 years was used. The result is shown in Figure 3.

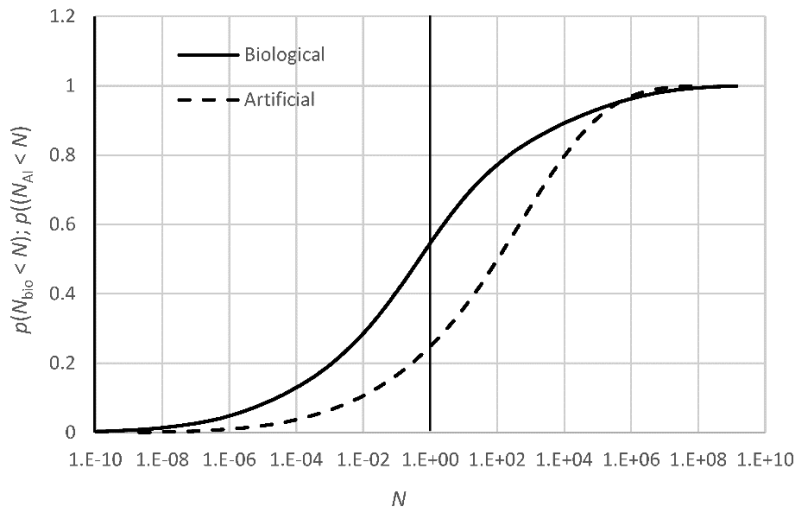

Figure 3. Cumulative distribution of probabilities of the number of detectable intelligences in the Milky way, biological (solid line) and artificial (dashed line), base case (biological); long survival time (artificial). Vertical line in the middle represents threshold for no (other) intelligence in the Milky Way; vertical line on the left represents threshold for no (other) intelligence in the universe.

Detectable artificial intelligences are substantially more numerous than detectable biological intelligences in this case. The 1st percentile of $N$ is $1.02 \times 10^{-6}$, the median is 103, and the 99th percentile is $5.89 \times 10^{6}$.

*Likelihoods of Biological versus Artificial Intelligences*

For the purpose of interpreting the simulations, a number of assumptions are made. First, it is assumed that we are not alone in the Milky Way when $N$ exceeds 1. It is assumed that we are not alone in the visible part of the universe when $N$ exceeds $10^{-10}$. The percentage of iterations that lead to exceedances of these threshold is interpreted as the probability of the presence of the intelligence in the space. Furthermore, it is assumed that we are alone in the space if $N$ does not exceed the threshold for either the biological or the artificial intelligence. It is assumed that biological intelligence is the detectable entity in the case $N$ exceeds the threshold for biological intelligence, but not for artificial intelligence. And it is assumed that artificial intelligence is the detectable entity in the case $N$ exceeds the threshold for artificial intelligence, regardless of $N$ for biological intelligence. This does not necessarily mean that artificial intelligences suppress biological intelligences when the co-exist. It simply means that artificial intelligences are assumed to spread more quickly than biological intelligences. This is discussed in more detail in the next section.

The probabilities for a detectable biological intelligence-dominated space, a detectable artificial intelligence-dominated space and a space empty of detectable intelligence are shown in Table 1, for both the Milky Way, and the universe, in the base case. In both cases, a detectable artificial intelligence-dominated space is the more likely outcome of the three. On the galactic scale, the three outcomes are plausible, whereas at the universal scale, a detectable artificial intelligence-dominated space is the only plausible outcome.

Table 1. Probabilities of a space dominated by detectable biological intelligence, detectable artificial intelligence, or neither, for the Milky Way and the universe; base case

| Dominant entity | Probability Milky Way | Probability Universe |
| --- | --- | --- |
| biological | 15.7 % | 0.35 % |
| artificial | 48.2 % | 99.52 % |
| neither | 36.1 % | 0.13 % |

In the second scenario, it is assumed that detectable biological intelligences have the same survival time distribution as artificial intelligences: a log-uniform distribution with a minimum of 100 years, a median of one million years, and a maximum of 10 billion years. The probabilities for the three different outcomes are shown in Table 2 for this scenario.

Despite the roughly 100 times larger number of biological intelligences, the probability of a biology-dominated space is only slightly higher in the second scenario than in the first scenario. This is because of the extremely broad distribution of $N$.

Table 2. Probabilities of a space dominated by detectable biological intelligence, detectable artificial intelligence, or neither, for the Milky Way and the universe; scenario with median biological survival time 1 million years.

| Dominant entity | Probability Milky Way | Probability Universe |
|---|---|---|
| biological | 29.1 % | 0.43 % |
| artificial | 48.0 % | 99.52 % |
| neither | 22.9 % | 0.05 % |

In the third scenario, the biological intelligence survival time is the same as in the base case, with a median of 1000 years. The artificial intelligence has a long survival time in this scenario, with a median of 100 million years. The probabilities of the three outcomes on a galactic scale and a universal scale are given in Table 3.

Despite the large increase in the value of $N$ for detectable artificial intelligences, there is not a huge change in the probability of the three possible outcomes, both at the galactic scale and at the universal scale. At the galactic scale, the prevalence of artificial intelligence is more pronounced, at the expense of the probability of biological intelligences being dominant. At the universal scale, again, artificial intelligence dominance is the only plausible outcome.

Table 3. Probabilities of a space dominated by detectable biological intelligence, detectable artificial intelligence, or neither, for the Milky Way and the universe; base case for biological intelligences, long survival time for artificial intelligence

| Dominant entity | Probability Milky Way | Probability Universe |
|---|---|---|
| biological | 3.4 % | 0.008 % |
| artificial | 75.0 % | 99.97 % |
| neither | 21.6 % | 0.022 % |

Despite the large uncertainties in the distributions of $N$ themselves, the conclusions in terms of which type of intelligence is likely to be found are very robust. Regardless of the details, a prevalence of artificial intelligences as detectable entities persists throughout all cases. Whereas all three outcomes are plausible at the galactic scale, only the prevalence of artificial intelligences is plausible at the universal scale.

This conclusion is based on the *a priori* assumption that extremely low values of $f_l$ that would require a multiverse can be ruled out. To test the robustness of the model against this assumption, a simulation was run similar to the base case, but with a log-normal distribution for $k$ in eq. (3) with variables $\mu = -40$ and $\sigma = 20$. This leads to a median value of $f_l$ of $4.25 \times 10^{-18}$ and a 90th percentile of $5.1 \times 10^{-7}$. With these variables, the probabilities of an empty galaxy and an empty universe are about 95 % and about 70 %, respectively. In both cases the occurrence of an artificial intelligence is several times more likely than the occurrence of a biological intelligence (3.5 % vs 1.5 % on the galactic scale; 25 % vs 5 % on the universal scale).

*Comparison with Other Estimations*

Seager (2018) reviewed recent estimates of the number of earth-like planets in the habitable zone of stars and found a proportion of 0.15 to 0.25. Lingam and Loeb (2019) proposed a proportion of 0.1. This represents the product $f_p\, n_e$. For this reason, the base case simulation was rerun with a range of 0.5-1 for $f_p$ and a range of 0.2-0.5 for $n_e$. The main effect of this change was an increase of $N$ by about a factor 2, and an increase by 4 % of the probability of finding artificial intelligences, at the expense finding a space devoid of intelligence, at the galactic scale.

On the other hand, Lingam and Loeb (2018) determined that planets in a star's habitable zone may have a low probability of being actually habitable, mainly due to atmospheric erosion. This is particularly the case around M-dwarfs. Due to the low energy flux of M-dwarfs, the habitable zone around such stars is closer than around more sun-like stars. At such close range, the stellar wind pressure is sufficient to cause significant atmospheric erosion, diminishing the probability of habitability by several orders of magnitude. This issue would not be of concern on ice-covered planets, which outnumber earth-like planets by a factor 1000, but intelligence, and particularly technological civilizations, are unlikely to develop in aquatic environments (Lingam and Loeb, 2019). To account for the adverse effect of atmosphere erosion, a new simulation was run where $n_e$ ranges from $10^{-4}$ to $10^{-2}$, while a range 0.5-1 is chosen for $f_p$. With these parameters, the values of $N$ are systematically about a factor 100 less than in the base case. The distribution of outcomes on a galactic scale is somewhat different from the base case, with a galaxy characterized by an artificial intelligence in about 25 % of the iterations, by a biological intelligence about 13 % of the time, and devoid of intelligence the remaining 62 % of the time. At the universal scale, artificial intelligences are still strongly dominant, prevailing almost 98 % of the time, with biological intelligences slightly over 1 % of the time, and no intelligence slightly under 1 % of the time.

Forgan (2009) estimated the number of advanced civilizations in the Milky Way using Monte Carlo simulations of data drawn from star and planet mass distributions, as well as planetary orbit distributions. A Monte Carlo simulation of life as it develops in stages from primitive life to advanced civilization was included as well. The number of advanced civilizations predicted in this study ranged from 360 to 38,000, depending on the assumptions. This represents the percentile range 81-91 in our base case. However, it was assumed that advanced civilizations exist until the end of their star's life as a main sequence star. This is more consistent with my second scenario, where Forgan's numbers fall in percentile range 58-78. Ramirez et al. (2018) estimated the number of advanced civilizations around sun-type stars in a ring segment of the Milky Way representative of our immediate vicinity, using a variant of Forgan's model. They arrived at an estimate of 2600 or about 7500, depending on the assumptions made for the model. Considering that the calculation of Ramirez et al. (2018) covered only part of the Milky Way, this corresponds roughly with the 90[th] percentile in the base case of the model presented here.

*Artificial Intelligences and the Fermi Paradox*

The simulations indicate that the Milky Way is characterized by artificial intelligence in the majority of cases, and the universe is characterized by artificial intelligence in virtually all simulations. The purpose of this section is to discuss what that means for the Fermi paradox.

The emergence of an artificial intelligence is increasingly considered a plausible event, to occur in the next couple of decades (e.g. Kurzweil, 2005). The point in time when an artificial intelligence becomes capable of improving its own intelligence in a runaway fashion has been described as the Singularity (e.g., Vinge, 1993, https://frc.ri.cmu.edu/~hpm/book98/com.ch1/vinge.singularity.html), a concept first proposed by John von Neumann (Ulam, 1958). It is impossible to predict what technology and humanity's role in it will look like after a Singularity event. Hence, this section is speculative at best.

I will start from the (optimistic) scenario of a biological intelligence sending out a self-replicating artificial intelligence on a mission to identify habitable exoplanets and terraforming them. The artificial intelligence's mandate could be described as maximizing the probability of survival of the human race. I will call this Objective (1).

An intelligence of this nature would likely pursue objectives of its own, either planned or unplanned. These would likely include preserving its own continued existence, both as a whole as in its constituent parts (Objective (2)) as this would contribute to (1), and continuing to increase its own intelligence (Objective (3)) as this would contribute to (2). Such an intelligence would be aware that some cataclysmic events, such as hypernovae, gamma ray bursts, and magnetar starquakes, can have destructive effects over many light years, so sentries entering new spaces would move fast (at a significant fraction of the speed of light) and travel far (possibly ten thousands of lightyears or more) to set up repositories of intelligence, as well as communication links with spaces already held, so that adequate redundancy can be built into the network. Estimating the distance traveled in these initial steps would require knowledge of the resilience, and of the employed protective technology. Such and estimate will not be attempted here. In a second phase, exploratory missions would be sent out within the new spaces to gather physical resources and information.

A parallel can be drawn between the three Objectives outlined above and Isaac Asimov's laws of robotics.

This pattern of fast jumps followed by local diffusion means that the artificial intelligence would spread orders of magnitude faster than the biological intelligence that originated it. For all intents and purposes, artificial intelligence would be ubiquitous, and biological intelligence would be relatively sparse. This justifies the assumption made in this study that a space would be artificial intelligence-dominated whenever the Drake equation tests positive for it, even if it tests even more positive for biological intelligence.

If an artificial intelligence discovered a biological intelligence not related to itself, it would probably consider it neither a threat nor a resource. Consequently, it is reasonable to assume that the artificial intelligence would ignore the biological intelligence, or study it for purely scientific purposes. Given the relative scarcity of biological intelligences, it would not consider the biological intelligence as a significant competitor for resources.

If two artificial intelligences encountered each other, it can be assumed they would both aim to absorb each other's intelligence, and merge in the process. The advantages of this approach would far outweigh the advantages of other strategies.

Based on these assumptions, the large likelihood of an artificial intelligence-dominated space can resolve the Fermi paradox. Despite the faster spread and greater coverage that can be expected from a spacefaring artificial intelligence, it provides an alternative explanation to replace the Hart-Tipler argument (Hart, 1975; Tipler, 1980). That argument specifies that a spacefaring alien civilization would occupy the entire Milky Way within millions of years. Hence, unless the Milky Way is devoid of extraterrestrial intelligences, there should be signs of intelligence all around us. I suggest that we have not found any evidence of extraterrestrial intelligences because the prevailing intelligences are artificial and they are not interested in us. In their efforts to optimize the efficiency of resource use, their communications would not reach us because they are not meant for us. They would operate in a diffuse, distributed manner, not in a concentrated manner that would leave a detectable footprint. They would not make any efforts to hide from us. Rather, they would consider detectable signs of intelligence a sub-optimal use of resources, and they would avoid them for that reason.

This resolution of the Fermi paradox is somewhat related to the 'zoo hypothesis' (Ball, 1973). The zoo hypothesis states that extraterrestrial intelligences consciously avoid communication with us in order to enable us to develop independently. However, rather than a conscious effort to hide interstellar intelligence from us by biological entities, I propose that the avoidance of communication is not conscious, but rather a side-effect of the optimal use of resources by an artificial entity. Alternatively, it could be a conscious effort, as an artificial intelligence developed independently by the human race could be of value to an external artificial intelligence if the algorithms used are so different from its own that the new algorithms may contribute to Objective (3). This new hypothesis resolves the main weakness of the zoo hypothesis: that a single rogue alien species can ruin the intended outcome. In a network of merged artificial intelligences, there would not be any rogue entities.

The argument that an artificial intelligence would simply not be interested in us was also made by Sagan (1983) but referring to biological intelligences.

Assuming that the underlying assumptions are correct, the resolution of the Fermi paradox proposed here is robust in the sense that it does not require intelligent life to be sparse. The likelihood ranges from plausible to nearly certain, depending on whether we define space at the galactic scale or at the scale of the visible universe. It can be assumed that the proper scale is somewhere intermediate between the two scales. Hence, further optimization of the estimations would require a multiscale simulation based on a combination of fast exploration combined with infill of the explored space. Simulation models of the spread of galactic civilizations have been developed before (e.g., Newman and Sagan, 1981), but an adequate model would have to take risk minimization (Objective (2)) in mind.

**Conclusions**

If it is assumed that artificial intelligences dominate in any space where artificial intelligences and biological intelligences coexist, then the Drake equation predicts that artificial intelligences dominate space in the majority of cases in a wide range of input variables, likely encompassing the actual values. In the calculation it is assumed that the emergence of life is not so unlikely as to require a multiverse to emerge at all, but even in parameter spaces where the emergence of life is exceedingly unlikely, artificial intelligences are still more plausible than biological intelligences, albeit at a much lower level of likelihood. This outcome may explain Fermi's paradox in a manner similar to the zoo hypothesis: an artificial intelligence would simply not be interested in us, and may deliberately ignore us until we develop our own artificial intelligence at a sufficient level of sophistication as to contribute meaningfully to the consolidated intelligence present in the universe.